\input harvmac
\input amssym.def
\input amssym.tex
\input epsf.tex

\font\male=cmr9
\font\tfont=cmbx12 scaled\magstep1 
\def\dia{~~$\diamondsuit$}
\hfuzz 20pt
\def\nl{\hfil\break}

\def\tch{\widetilde \ch}
\def\sgn{{\rm sign}}
\def\ch{{\rm ch}}
\def\l{\lambda}
\def\half{{\textstyle{1\over 2}}}
\def\oom{{\textstyle{1\over m}}}
\def\jkm{{\textstyle{jk\over m}}}
\def\tom{{\textstyle{2\over m}}}
\def\qtr{{\textstyle{1\over 4}}}
\def\oei{{\textstyle{1\over 8}}}
\def\oec{{\textstyle{c\over 8}}}
\def\oes{{\textstyle{1\over 16m}}}

\def\d{\delta} \def\ls{{\rm l.s.}}
\def\eps{\epsilon}
\def\bbc{{C\kern-6.5pt I}}
\def\bac{{C\kern-5.5pt I}}
\def\bbn{I\!\!N}
\def\bbz{Z\!\!\!Z}

\def\bG{{\bar G}}
\def\ca{{\cal A}}  
  
\def\cg{{\cal G}} \def\cH{{\cal H}}

\def\cp{{\cal P}} 
\def\ct{{\cal T}}


\lref\Ade{M. Ademollo et al, Phys. Lett. {\bf B62} (1976) 105-110;
Nucl. Phys. {\bf B111} (1976) 77-110.}

\lref\BFK{W. Boucher, D. Friedan and A. Kent, Phys. Lett. {\bf B172} (1986) 316-322.}

\lref\Dosh{V.K. Dobrev, in:  Proc. XIII
International Conference on Differential-Geometric Methods in
Theoretical Physics, Shumen (1984), eds. H.D. Doebner and T.D.
Palev, (World Sci., Singapore, 1986) pp. 348-370.}

\lref\Dob{V.K. Dobrev, Phys. Lett. {\bf B186} (1987) 43-51.}

\lref\RC{A. Rocha-Caridi, in: "Vertex Operators in Mathematics and Physics",
eds. J. Lepowsky, S. Mandelstam and I. Singer (Springer-Verlag, M,Y., 1985) p. 451.}

\lref\Doa{V.K. Dobrev, Suppl. Rend. Circ. Mat. Palermo, Serie II, Numero 14
(1987) 25-42.}

\lref\Dolmp{V.K. Dobrev, Lett. Math. Phys. {\bf 11} (1986) 225-234.}

\lref\FF{B.L. Feigin  and D.B. Fuchs,
Funkts. Anal. Prilozh. {\bf 17} (3) (1983) 91-92,\
English translation: Funct. Anal. Appl. {\bf 17} (1983) 241-242;
~Lecture Notes in Math. vol. 1060 (1984) pp. 230-245.}

\lref\Dz{M. Doerrzapf,   Nucl. Phys. {\bf B529} (1998) 639-655.}

\lref\Mat{Y. Matsuo, Prog. Theor. Phys. {\bf 77} (1987) 793-797.}

\lref\Kir{E. Kiritsis, Int. J. Mod. Phys. {\bf A3} (1988) 1871-1906.}


\centerline{{\tfont Characters of the Unitarizable Highest Weight
Modules}}
\vskip
2truemm \centerline{{\tfont over the N=2 Superconformal Algebras}
\foot{This is a slightly extended version  of an Encyclopedia entry.}}

\vskip 0.5cm

\centerline{{\bf V.K. Dobrev}}
\centerline{\male Institute for Nuclear Research and Nuclear Energy}
\centerline{\male Bulgarian Academy of Sciences, Sofia, Bulgaria}

\vskip 0.5cm

  The ~{\it N=2 superconformal} (or {\it  super-Virasoro}) {\it algebras}~ in two dimensions
are three complex Lie superalgebras: the ~{\it Neveu-Schwarz}~ superalgebra \Ade,
the ~{\it Ramond}~ superalgebra \Ade, the ~{\it twisted}~ superalgebra \BFK,
which are denoted  as $\ca$, $\cp$, $\ct$, resp., or $\cg$ when
a statement holds for all three superalgebras. They have
the following nontrivial super-Lie brackets :
\eqna\virnt
$$\eqalignno{
[L_m,L_n] ~=&~ (m-n)\ L_{m+n} ~+~  \qtr\,z\,\left( m^3-m \right)\, \d_{m,-n}\
 &\virnt a\cr
[L_m,G^j_n] ~=&~ (\half m-n)\ G^j_{m+n}  \ , \quad j=1,2
 &\virnt b\cr
[L_m,Y_n] ~=&~ -n\, Y_{m+n}\ ,\qquad
[Y_m,Y_n]  ~=~ z\,m\, \d_{m,-n}\    &\virnt c\cr
[Y_m,G^j_n] ~=&~ i\,\eps^{jk}\, G^k_{m+n}  \ , \quad \eps^{jk} = \pmatrix{0 &1\cr -1 & 0}
 &\virnt d\cr
[G^j_m,G^k_n]_+ ~=&~ 2\,\d^{jk}\, L_{m+n} ~+~ i\,\eps^{jk}\,(m-n)\ Y_{m+n}
~+~ z\,\left( m^2-\qtr \right)\,\d^{jk}\, \d_{m,-n}\
  &\virnt e\cr}$$
where $m\in\bbz$  in $L_m$ for all superalgebras; ~$m\in\bbz$   in $Y_m$ and
$m\in\half +\bbz$ in $G^j_m$   for $\ca$;
~$m\in\bbz$   in $Y_m$ and   $G^j_m$   for  $\cp$;
~$m\in\bbz$   in $G^1_m$ and $m\in\half +\bbz$ in $Y_m$ and $G^2_m$    for  $\ct$.\nl
\indent     The standard triangular decomposition of $\cg$ is:
\eqna\tria
\eqna\tric \eqna\trid
$$\eqalignno{
&\cg = \cg_+ \oplus \cH \oplus \cg_- &\tria {}\cr
\cH ~=&~ \ls \{z\,, L_0\, ,\ Y_0\,\}\ \quad {\rm for} ~\ca, \cp &\tric a\cr
~=&~ \ls \{z\,, L_0\, ,\ G^1_0\,\}\ \quad {\rm for} ~\ct,
\quad (G^1_0)^2  = L_0 - z/8 &\tric b\cr
\cg_+ ~=&~ \ls \{ L_m\, , m>0, ~Y_n\,, n>0, ~G^j_p\, ~p>0 \} ~\oplus~
\ls \{ {\bar G}_0 \}_\cp &\trid a\cr
\cg_- ~=&~ \ls \{ L_m\, , m<0, ~Y_n\,, n<0, ~G^j_p\, ~p<0 \} ~\oplus~
\ls \{ {G}_0 \}_\cp &\trid b\cr}$$
where the generators $G_0\,,{\bar G}_0$ which appear for  $\cp$ in \trid{} are the zero modes
of:
\eqn\zerg{ G_n ~=~ \half\big( G^1_n + iG^2_n)\ , \qquad
\bG_n ~=~ \half\big( G^1_n - iG^2_n) }

A highest weight module (HWM) over $\cg$  is characterized by its highest weight $\l\in\cH^*$
and highest weight vector $v_0$ so that $X\,v_0 = 0$, for $X\in\cg_+\,$, ~$H\,v_0 = \l(H)v_0$ for $H\in\cH$.
Denote ~$\l(L_0)= h$, ~$\l(z) = c$, ~$\l(Y_0) = q$. [Note that interchanging $G_0$ and $\bG_0$
in \trid{} means to pass from $P^+$ to $P^-$ modules in the terminology of \BFK.]
The largest HWM with these properties is the Verma module $V^\l = V^{h,c,q}$ ($= V^{h,c}$ for $\ct$),
which is isomorphic to $U(\cg_-)v_0\,$, where $U(\cg_-)$ denotes the universal enveloping algebra
of $\cg_-\,$. Denote by $L^\l$ (resp. $L^{h,c,q}$, $L^{h,c}$) the factor-module $V^\l/I^\l$,
where $I^\l$ is the maximal proper submodule of $V^\l$. Then every irreducible HWM over $\cg$
is isomorphic to some $L^\l$.

A Verma module $V^{h,c,q}$ ($V^{h,c}$) over $\cg$ is reducible if and only if \BFK:
 \eqna\reda\eqna\redp\eqna\redt
$$\eqalignno{
&f^A_{r,s} \equiv 2h(c-1) -q^2 -\qtr(c-1)^2 + \qtr [(c-1)r+s]^2 = 0,
~~~{\rm for~ some} ~r\in\bbn, s\in 2\bbn,\quad  &\reda a\cr
&or ~~~g^A_n \equiv 2h- 2nq + (c-1) (n^2-\qtr) = 0, ~~~{\rm for~ some} ~ n\in\half+\bbz,
\quad {\rm for} ~\ca\ ;&\reda b\cr
&f^P_{r,s} \equiv 2(c-1)(h-\oei) -q^2  + \qtr [(c-1)r+s]^2 = 0,
~~~{\rm for~ some} ~r\in\bbn, s\in 2\bbn,\quad  &\redp a\cr
&or ~~~g^P_n \equiv 2h- 2nq + (c-1) (n^2-\qtr) -\qtr = 0, ~~~{\rm for~ some} ~ n\in\bbz,
\quad {\rm for} ~\cp\ ;&\redp b\cr
&f^T_{r,s} \equiv 2(c-1)(h-\oei)  + \qtr [(c-1)r+s]^2 = 0,
~~~{\rm for~ some} ~r\in\bbn, s\in 2\bbn-1, \quad {\rm for} ~\ct
&\redt {}\cr
}$$
\indent The necessary conditions for the unitarity of $L^{h,c,q}$ ($L^{h,c}$) are \BFK:
\eqna\unia\eqna\unip\eqna\unit
$$\eqalignno{
&{\rm case}~A_3\,: \quad c\geq 1, ~g^A_n\geq 0, ~~~{\rm for~all}~ n\in \half +\bbz \ ; &\unia a\cr
&{\rm case}~A_2\,: \quad c\geq 1, ~f^A_{1,2}\geq 0, ~g^A_n= 0,
 ~g^A_{n+{\rm sign}(n)} \leq 0, ~~~{\rm for~some}~ n\in \half +\bbz \ ; &\unia b\cr
&{\rm case}~A_0\,: \quad  c<1, ~c =1-\tom , ~~h = \oom(jk-\qtr) , ~~q = \oom(j-k)  , \cr
&\qquad\qquad ~~~{\rm for}~~ m\in 1+\bbn, ~~j,k \in \half +\bbz, ~~~
 0< j,k,j+k \leq m-1\ ;&\unia c\cr
&{\rm case}~P_3\,: \quad c\geq 1, ~g^P_n\geq 0, ~~~{\rm for~all}~ n\in \bbz \ ; &\unip a\cr
&{\rm case}~P_2\,: \quad c\geq 1, ~f^P_{1,2}\geq 0, ~g^P_n= 0,
 ~g^P_{n+{\rm sign}(n)}< 0, ~~~{\rm for~some}~ n\in \bbz \ , &\unip b\cr
&\qquad\qquad\qquad {\rm sign}(0)=\pm1 ~{\rm for}~ P^\pm~;   \cr
&{\rm case}~P_0\,: \quad  c<1, ~c =1-\tom , ~~h = \oei c +\jkm  , ~~q = \pm\oom(j-k)  , \cr
&\qquad\qquad ~~~{\rm for}~~ m\in 1+\bbn, ~~j,k \in \bbz, ~~~
 0\leq j-1,k,j+k \leq m-1\ ;&\unip c\cr
&{\rm case}~T_2\,: \quad c\geq 1, ~~ h\geq \oei c\ ; &\unit a\cr
&{\rm case}~T_0\,: \quad  c<1, ~c =1-\tom , ~~h = \oei c + \oes (m-2r)^2,  \cr
&\qquad\qquad ~~~{\rm for}~~ m\in 1+\bbn, ~~ r \in \bbn, ~~~
 1\leq r  \leq \half m\ ;&\unit b\cr
 }$$
\indent  Further  write $V^{h,c,(q)}, L^{h,c,(q)}$ in the cases when a statement holds for
$V^{h,c,q}, L^{h,c,q}$ over $\ca,\cp$ as written and for $V^{h,c}, L^{h,c}$
over $\ct$ after deleting q and all related quantities.

The weight decomposition of $V^{h,c,(q)}$~ is:
\eqna\deco
$$\eqalignno{
V^{h,c,(q)} ~=&~ \mathop{\oplus}\limits_{n,(m)} V_{n,(m)}^{h,c,(q)} &\deco a\cr
V_{n,(m)}^{h,c,(q)} ~=&~ \{\ v\in V^{h,c,(q)} ~\vert~
L_0\, v = (h+n)v , ~~{\rm for}~ \cg \ , \quad
Y_0\, v = (q+m)v , ~~{\rm for}~ \ca\,,\cp \ \}\qquad &\deco b\cr}$$
where the ranges of $n,m$  in \deco{} are:
\eqna\decs
$$\eqalignno{
&n \in \half \bbz_+\ , ~~~m \in 2n + 2\bbz\,, ~~~ \vert m\vert \leq
\sqrt{2n}\,, ~~~{\rm for}~ \ca &\decs a\cr
&n \in  \bbz_+\ , ~~~m \in \bbz\,, ~~~ \vert
\half (1- \sqrt{8n+1}) \leq m \leq  \half (1- \sqrt{8n+1})\,,
  ~~~{\rm for}~ \cp &\decs b\cr
&n \in \half \bbz_+\ , ~~~{\rm for}~ \ct &\decs c\cr }$$
$n$ is called the level of $V_{n,(m)}^{h,c,(q)}$, ~$m$ - its relative charge.

Then the character of $V^{h,c,(q)}$ may be defined as follows \BFK:
\eqna\chav\eqna\chaw
$$\eqalignno{ \ch\, V^{h,c,q} ~=&~ \mathop{\sum}\limits_{n,m}\,
(\dim\ V_{n,m}^{h,c,q})\, x^{h+n}\, y^{q+m} ~=~
\mathop{\sum}\limits_{n,m}
P(n,m)\, x^{h+n}\, y^{q+m} ~=~ x^h\, y^q\, \psi(x,y)\qquad\quad
 &\chav a\cr
\ch\, V^{h,c} ~=&~ \mathop{\sum}\limits_{n}\,
(\dim\ V_{n}^{h,c})\, x^{h+n}  ~=~
\mathop{\sum}\limits_{n}
P_T(n)\, x^{h+n}  ~=~ x^h\,  \psi_T(x)\
, &\chav b\cr
\psi_A(x,y) \equiv& \sum_{n,m}P_A(n,m)\, x^{n}\, y^{m} = \prod_{k\in\bbn}
{(1+x^{k-1/2}y) (1+x^{k-1/2}y^{-1})\over (1-x^k)^2} &\chaw a\cr
\psi_P(x,y) \equiv& \sum_{n,m}P_P(n,m)\, x^{n}\, y^{m-1/2} = (y^{1/2} + y^{-1/2})
\prod_{k\in\bbn}
{(1+x^{k}y) (1+x^{k}y^{-1})\over (1-x^k)^2} \quad &\chaw b\cr
\psi_T(x) \equiv& \sum_{n}P_T(n)\, x^{n} = \prod_{k\in\bbn}
{(1+x^{k}) (1+x^{k-1/2})\over (1-x^k) (1-x^{k-1/2}) } &\chaw c\cr
}$$
(for $P^-$ representations one should write $y^{m+1/2}$ instead of $y^{m-1/2}$ \BFK).

{\bf Proposition 1:} \BFK,\Dob ~~~The character formulae for the unitary cases $A_3$, ($P_3$),
with either $c>1$ and $g_n>0$, $\forall n\in\half+\bbz$, ($\forall n\in\bbz$),
 or ~$c=1$, and cases $T_2$ are given by:
\eqna\chauir
$$\eqalignno{ \ch\, L^{h,c,q} ~=&~  \ch\, V^{h,c,q} &\chauir a\cr
\ch\, L^{h,c} ~=&~  \ch\, V^{h,c} \ , ~~~h\neq \oec \ ,\qquad
\ch\, L^{\oec,c} ~=~ \half\ \ch\, V^{\oec,c}
&\chauir b\cr}$$
Note that the Verma modules involved are irreducible except in the last case, where
~$V^{\oec,c} = I^{\oec,c} \oplus V^{\oec,c}/ I^{\oec,c}\ $,
~~~$I^{\oec,c} \cong V^{\oec,c}/ I^{\oec,c}\ $.\dia

{\bf Proposition 2:} \Dob ~~~The character formulae for the unitary cases $A_3$, ($P_3$),
with $c>1$, ~$q/(c-1) = n_0\in \half+\bbz$, ($n_0\in \bbz$),
and $g_{n_0}=0$,  and for the cases $A_2$, ($P_2$), with $f_{1,2}>0$,
are given by:
\eqn\chaut{ \ch\, L^{h,c,q} ~=~ {\widetilde \ch}_n \ V^{h,c,q} ~\equiv~
{1\over (1+ x^{|n|}y^{{\rm sign}(n)}) }\ \ch\, V^{h,c,q}\ }
where for $A_3$, $P_3$, ~$n=n_0$,
 and for $A_2$, $P_2$, ~$n$ is such that ~$g_n=0$,
~$g^A_{n+{\rm sign}(n)} <0$.\nl {\it Proof:}~~~ Actually, the
Proposition holds in a more general situation beyond the unitary
cases, namely, when for a fixed $V^{h,c,q}$  \reda{b}, (\redp{b})
holds for some $n$, possibly also for some $n'$ such that ~${\rm
sign}(n)={\rm sign}(n')$ and ~$|n'|>|n|$, and \reda{a}, (\redp{a})
does not hold for any $r,s$. [In the statement of Proposition 2 the
additional reducibility appears in the cases $A_2$, ($P_2$) when
~$2q(c-1) \in\bbz$, then $n'=M-n$, $M \equiv 2q(c-1)$  and
~$g_{M-n}=0$.] In this situation there is a singular vector $v^s_n$
and possibly a singular vector $v^s_{n'}\,$, however, the latter
(when existing) is a descendant of $v^s_n\,$. Thus, there is the
following embedding diagram: \eqn\emb{ V^{h,c,q} ~\longrightarrow~
V^{h+|n|,c,q+{\rm sign}(n)}  } where is used the convention that the
arrow points ~{\it to}~ the embedded module. This embedding has a
kernel, since there is an infinite chain of embeddings of Verma
modules: \eqn\embs{\cdots ~\longrightarrow~ V_t  ~\longrightarrow~
V_{t+1}  ~\longrightarrow~ \cdots } where $V_t \equiv
V^{h+t|n|,c,q+t{\rm sign}(n)}$, $t\in\bbz$. Using the   Grassmannian
properties of the odd generators one can show that this chain of
embedding maps is exact. Due to the kernel one has: \eqn\proo{ \ch\,
L^{h,c,q} ~=~ \ch\, V^{h,c,q} ~-~ \widetilde{\ch}_n \,
V^{h+|n|,c,q+{\rm sign}(n)} ~=~ {\widetilde \ch}_n \, V^{h,c,q} }

{\bf Proposition 3:} \Dob ~~~The character formulae for the unitary cases $A_2$, $P_2$,
with ~$f_{1,2}=0$  is given by:
\eqn\chautt{
\ch\, L^{h,c,q} ~=~  { (1-x)
\over (1+ x^{|n|}y^{{\rm sign}(n)})\,
(1+ x^{|n|+1}y^{{\rm sign}(n)})
}\ \ch\, V^{h,c,q}\ }
{\it Proof:}~~~ The character relevant structure of $V^{h,c,q}$ is given by the embedding diagram:
\eqn\embt{ \epsfbox{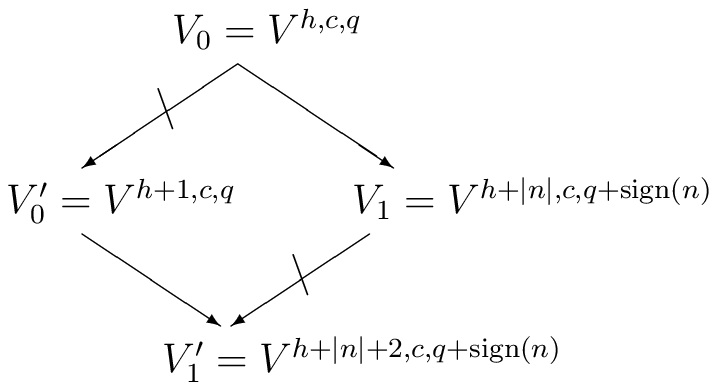}}
where the dashed arrows denote even embeddings, $V_0$ is reducible v.r.t. $g_n=0=f_{1,2}$
from the statement; then  (with $\mu = 0$ for $\ca$, $\mu = 1$ for $\cp$)~:
\eqn\valu{
h = \oei (c-1)(2n+\eps)^2 + n\eps + \oei\mu\ ,
\qquad q = \half (c-1)(2n+\eps)\ , \qquad \eps \equiv \sgn(n) }
The other reducibilities relevant for the structure are: $V_1$ w.r.t. $g_n=0=f_{1,2}\,$,
~$V'_0\,$ and $V'_1\,$  w.r.t. $g_{n+\sgn(n)}=0$. Thus for the
character formula follows:
\eqn\calch{\ch\, L^{h,c,q} ~=~ \ch\, V^{h,c,q} ~-~ \ch\, V^{h+1,c,q}
~-~ \widetilde{\ch}_n \, V^{h+|n|,c,q+{\rm sign}(n)} ~+~
\widetilde{\ch}_{n+\sgn(n)} \, V^{h+|n|+2,c,q+{\rm sign}(n)} }
which after substituting the definitions gives \chautt.\dia

{\bf Proposition 4:} \Dob,\Mat,\Kir ~~~The character formulae for the unitary cases
$A_0$, $P^\pm_0$, is given by:
\eqn\chauz{\eqalign{
\ch\, L_{m,j,k}(x,y) ~=&~ \sum_{n\in\bbz_+} x^{mn^2 + (j+k)n} \Big\{\
1 - x^{(m-j-k)(2n+1)}\ + \Big. \cr
+&~ x^{mn+k}y \Big[\, {x^{2(m-j-k)(n+1)} \over 1+x^{mn+m-j}y} - {1\over 1+x^{mn+k}y}\, \Big]
\ +\cr
\Big. +&~ x^{mn+k}y^{-1}\Big[\, {x^{2(m-j-k)(n+1)} \over 1+x^{mn+m-k}y^{-1}}
- {1\over 1+x^{mn+j}y^{-1}}\, \Big]\Big\}\ \ch\, V_{m,j,k}(x,y) }}
where ~$L_{m,j,k} = L^{h,c,q}$, ~$V_{m,j,k} = V^{h,c,q}$, when $h,c,q$ are expressed through $m,j,k$
as in \unia{c}, \unip{c}.\nl
{\it Proof:} ~~~The structure of ~$V_0 \equiv V_{m,j,k}$~ is given by the following embedding diagram:
\eqn\embt{ \epsfbox{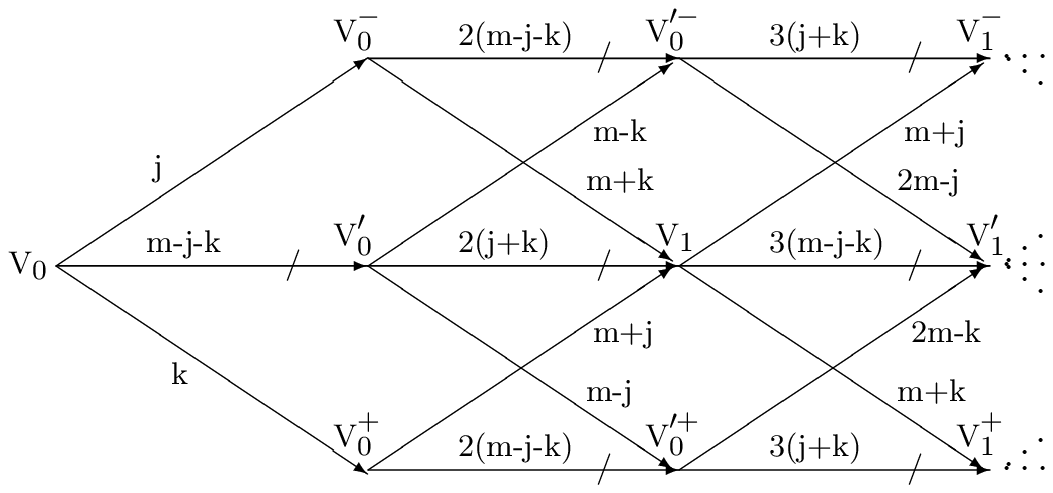}}
\eqn\param{\eqalign{ & V_n = V^{h+mn^2 +(k+j)n,c,q} \ ,
\quad  V'_n = V^{h+mn^2 +m(2n+1) -(k+j)(n+1),c,q} \ ,\cr
& V^+_n = V^{h+mn^2 +(m+k+j)n+k,c,q+1}\ , \quad
V'^+_n = V^{h+mn^2+ m(3n+2) -(k+j)(n+2)+k,c,q+1}\ , \cr
& V^-_n = V^{h+mn^2 +(m+k+j)n+j,c,q-1}\ , \quad
V'^-_n = V^{h+mn^2+ m(3n+2) -(k+j)(n+2)+j,c,q-1}\  \cr
}}
From this follows:
\eqn\chauzz{\eqalign{ \ch\, L_{m,j,k}(x,y) ~=~ \sum_{n\in\bbz_+} \Big[ \,
&\ch\, V_n ~-~ \ch\, V'_n ~-~ \tch_{mn+k}\, V^+_n ~-~ \tch_{mn+j}\, V^-_n ~+ \Big.\cr
\Big. &+~ \tch_{mn+m-j}\, V'^+_n ~+~ \tch_{mn+m-k}\, V'^-_n \, \Big] }}
which after substituting the definitions gives \chauz.\dia

{\bf Remark:}~~~ It should be stressed that diagram \embt{} is used   only
as representing the structure of the Verma module ~$V_{m,j,k}\,$. In particular,
later it was shown that each even embedding   between the Verma modules
$V_n$ and $V'_n$, $n=1,2,\ldots$, and between the Verma modules  $V'_n$ and $V_{n+1}$,
 $n=1,2,\ldots$, is generated by two uncharged fermionic singular vectors
   \Dz. However, this has no relevance for the character formulae.

{\bf Proposition 5:} \Dob,\Mat,\Kir ~~~  Let ~$V_{r,s}\,$, ~$r\in\bbn$, $s\in\bbn-1/2 $,
be the Verma module   $V^{h,c}$  with ~$h = h^T_{r,s} = [(tr-ms)^2 -t^2]/4mt + 1/8 =
h^T_{m-r,t-s}\,$, ~$c=1-2t/m$, ~$t,m\in\bbn$, ~$tr\leq ms$, ~$s<t<m$, ~$t,m$ have no common divisor.
Then the character formula for the corresponding irreducible quotient $L_{r,s}$ is given by:
\eqn\chauf{ \ch\, L_{r,s}(x) ~=~ \ch\, V_{r,s}(x) \sum_{j\in\bbz} x^{j(tmj+tr-ms)} (1 - x^{s(2mj+r)}) }
In particular, the character formula for the $T_0$ unitary cases $r\leq m/2$
 is obtained from \chauf\ by setting $t = 1$, $s = 1/2$.\nl
The Proof relies on the realization that the Verma modules
$V_{r,s}$   has exactly the structure of certain Virasoro and
$N=1$ super-Virasoro  (Neveu-Schwarz and Ramond) Verma modules
for which the character formulae were known
(see also the corresponding encyclopedia entry).\dia

\bigskip

\parskip=0pt
\listrefs

\end